*Article*

# A Hierarchical Security Event Correlation Model for Real-Time Threat Detection and Response

**Herbert Maosa [1], Karim Ouazzane [1] and Mohamed Chahine Ghanem [1,2]**

[1] Cyber Security Research Centre, London Metropolitan University, London N7 8DB, UK; h.maosa1@londonmet.ac.uk (H.M.); k.ouazzane@londonmet.ac.uk (K.O.)
[2] Department of Computer Science, University of Liverpool, Liverpool L69 3BX, UK
* Correspondence: ghanemm@staff.londonmet.ac.uk

**Abstract:** An intrusion detection system (IDS) perform postcompromise detection of security breaches whenever preventive measures such as firewalls do not avert an attack. However, these systems raise a vast number of alerts that must be analyzed and triaged by security analysts. This process is largely manual, tedious, and time-consuming. Alert correlation is a technique that reduces the number of intrusion alerts by aggregating alerts that are similar in some way. However, the correlation is performed outside the IDS through third-party systems and tools, after the IDS has already generated a high volume of alerts. These third-party systems add to the complexity of security operations. In this paper, we build on the highly researched area of alert and event correlation by developing a novel hierarchical event correlation model that promises to reduce the number of alerts issued by an intrusion detection system. This is achieved by correlating the events before the IDS classifies them. The proposed model takes the best features from similarity and graph-based correlation techniques to deliver an ensemble capability not possible by either approach separately. Further, we propose a correlation process for events rather than alerts as is the case in the current art. We further develop our own correlation and clustering algorithm which is tailor-made to the correlation and clustering of network event data. The model is implemented as a proof of concept with experiments run on standard intrusion detection sets. The correlation achieves an 87% data reduction through aggregation, producing nearly 21,000 clusters in about 30 s.

## 1. Introduction

Amoroso [1] defines alert correlation as "*. . . the interpretation, combination and analysis of information from all available sources about target system activity for purposes of intrusion detection and response.*" The goal of correlation is to reduce the number of events by aggregating and fusing those events that are related in some way. Additionally, through the correlation, it is possible to discover multistep and complex attacks whose information is carried in separate events. These events may also be generated by separate devices at various times and locations. From an intrusion detection perspective, correlation can be performed on either events or alerts. Alerts and events are related but fundamentally different concepts. The event is the record of what happened. The alert is a message that is raised when an interesting event has been encountered. The interestingness of an event depends on the application domain. From the Intrusion Detection Message Exchange Format (IDMEF) [2] specification, an alert is a message that is generated by a tool when it encounters an event that it was configured to look out for. Most of the active research is around alert correlation rather than event correlation. Event correlation is carried out before an IDS makes a detection decision. Alert correlation is performed outside the IDS after alerts have already been raised, by a third-party system, tool, or application. This paper deals with the correlation of the actual events before an analysis is performed by an IDS. We hypothesize that since the output of the correlation is a reduced data set due to aggregation, the input to the correlation component of an IDS is few correlated events. Consequently, the number of alerts should also be reduced.

### *1.1. Research Background*

Generally, correlation techniques are categorized as (1) similarity-based, (2) sequential, and (3) case−based [3]. Similarity−based techniques correlate two events if they are similar by comparing their attributes or time. Typically, some measure of distance is used to decide the similarity. Common distance functions in use are the Euclidean, Mahalanobis, Minkowski, and Manhattan distances. To determine if a new alert should be included in a cluster, its distance to the cluster is compared to some predetermined threshold. If the distance is above the threshold, it is correlated and included in the cluster. These techniques are known for their simplicity and performance in aggregation and reduction of the data set [4]. However, sequences in the events cannot be captured. Additionally, similarity-based techniques cannot be used to detect complex and multistep attacks. On the other hand, sequence-based techniques correlate events that have a causal relationship with each other. One event is considered the precondition while the other is the consequence. The strength of these techniques is their scalability and the ability to detect even previously unknown attacks. Case-based methods require the existence of a database of earlier cases which can be correlated with a new alert to make a classification and prediction. Several of these approaches have been used by researchers to address different correlation problems. What is clear is that there is no single approach that is universally applicable to every problem [3].

The problem being addressed by this paper can be characterized by the following requirements: (1) Correlation must be done on the events, rather than alerts. This approach is taken to deal with the issue of the high volume of alerts at the source, rather than after the fact. (2) The correlation must enable real-time attack detection so as to enable a fast response. This requirement means that approaches that require a long time or large data sets to build models and statistics do not apply, as those approaches will not satisfy the response times demanded by a real-time application.

Based on the problem defined and the approaches reviewed, this paper develops an event correlation model for real-time intrusion detection based on a hierarchy of two approaches: (A) a similarity correlation method is used to reduce the size of the data set and form clusters, and (B) graph correlation is applied on the clusters to discover cluster interconnections which can reveal unusual network traffic patterns that may indicate a possible attack. The target architecture is a distributed IDS where data collection components stream events to the correlation unit in real time. The proposed correlation model is described in detail in the rest of this paper.

### *1.2. Related Works*

This work is grounded in an area of expansive research in cybersecurity on alert correlation which aims at reducing the volume of alerts presented to cybersecurity analysts for ease of prioritization and response. The correlation process ingests alerts from several distributed IDSs, analyzes them, and constructs more compact reports that show a high-level view of the security status of a communications network as a whole [5]. Various research works have been carried out over the years on this topic across different application domains. In this section, we review only a select few of the available recent and pioneering research, focusing on those approaches we adopted in the listed contributions.

#### 1.2.1. Hierarchical Correlation

Different correlation techniques work best with different and specific types of correlation problems, data, and industries. Even within the same industry, each technique depends on the type of data available and the type of intrusion that is being pursued for detection. Hierarchical correlation is an approach that is based on the realization that no single approach is fit for all purposes. Using this approach, more than one technique is applied to the correlation problem with the reasoning that the combined approaches will achieve a far greater result than the individual approaches. The science, then, is to determine which combination of approaches works best for the problem at hand. There has been limited research on hierarchical correlation approaches. One of the pioneering works often referenced by researchers is the work by Cuppens and Miege [6], in which they performed a correlation of alerts from several distributed IDSs. The first level performed a clustering of alerts at a local level, while the second level correlated these clusters at the global level. However, they used the same correlation method at both levels of the hierarchy, namely, pre–post condition rules described in the LAMBDA language. Rule-based correlation, though simple and efficient, cannot be directly used to detect new attacks whose signatures may not yet be known. The hierarchy used in the approach by Cuppens relies on the physical spatial distribution of the IDSs and the correlation components.

In the model developed by Tian et al. [7], intrusion detection systems were distributed geographically in a network and their generated alerts were correlated at two levels. The first level performed local correlation using graphs to aggregate local alerts and discover local intrusion attempts. The generated attack graphs were then fed to a central correlation unit that further aggregated and correlated the graphs from the local correlation units to discover global intrusions. While this work achieved the detection of both local and global alerts, the authors also used the same technique for the correlation at both levels of the hierarchy. By using the same technique, the inherent weaknesses of the chosen approach were maintained throughout the correlation process flow. For example, graph-based correlations generate a lot of false positives [3], a limitation which will not necessarily be resolved even if the system is implemented in a hierarchy.

The system in [8] firstly aggregates alerts using a third-party system, **Splunk**, and then feeds the aggregated alerts to an automated program which uses text analysis to perform further correlation. The resulting correlated hyperalerts are then visualized using a sunburst diagram. This work increases the operational complexity of security operations because it increases the number of systems that need to be used for the analysis of security events. In addition to the IDS, the security analysts must deal with the security incident and event management (SIEM) system (Splunk) as well as the custom automated correlation application developed.

Natural language processing based on a combination of named-entity recognition (NER) and semantic similarity was used by [9]. They developed a correlation framework for cyberattack reconstruction. As a first stage, the alerts went through a reduction process, essentially aggregating them and constructing meta-alerts. Attributes of the aggregated alerts were then extracted using NER. If two alerts had at least a *threshold* number of similar attributes between them, they were then correlated based on semantic similarity. This approach was based on the analysis of textual data such as IDS alerts. It would be unsuitable for nontextual data, such as network traffic packet captures.

#### 1.2.2. Real-Time Correlation

Firstly, it is necessary to define real-time data. One such definition is given by [10], where real-time data are defined as "... information that is delivered immediately after collection. There is no delay in the timeliness of the information provided ...". Others have defined real-time data as data that are available when needed. Approaches that have been seen in research to optimize data collection for real-time performance include direct acquisition [11], event streaming [12–14], and the use of in-memory storage [15]. The key issue is to reduce the latency in the delivery of the event to the application that needs it. The latency is often caused by architectures that introduce too many hops such as aggregation points for the data, as well as access times due to the data being stored on disk or databases. In terms of real-time correlation and detection, the most common approach is to correlate all events that can fit into some defined time window. The challenge is to decide the optimal duration for the time window that will make the data available to the consumer application when needed. This question usually depends on some expert knowledge of the system and the type of data. If the time window is large, the time to analyze batched events and eventually react if there is an attack might be too long. This technique is seen in [16,17]. In [17], they follow the real-time algorithm from [17] for real-time processing of alerts. Incoming alerts are first grouped into batches that are sorted according to their creation time. Within the batch, the alerts are correlated according to a configurable time window, and each window is passed on to the analytics component as soon as it has been processed.

Zhang et al. [18] designed a detection system that operated in two phases. Initially, they mined frequent item sets for both normal and attack classes using nested sliding windows. Then, frequent item sets of incoming real-time network traffic flows were compared with these two classes. If the incoming flow was detected as anomalous, it was then passed on to the classification component, which used a combination of deep belief networks. Real-time processing was achieved by having the data processed from memory rather than from discs or databases. This significantly reduced the latency that is usually incurred due to disk access. The data are typically loaded into data structures in memory from where the correlation and analysis are run. Other researchers have used in-memory databases such as Oracle in-memory [19].

#### 1.2.3. Correlation Processes

The survey by Salah et al. [3] is one of the most referenced works on alert correlation techniques. They reviewed several works, and they summarized the alert correlation processes to propose a process model that comprised (1) alert processing, (2) alert reduction, (3) alert correlation, and (4) alert prioritization. Each of these process blocks expanded to other subprocesses. For example, preprocessing constituted normalization and feature construction, while alert reduction was made up of filtering and reduction. The issue with correlation processes is that until now, there is no standard approach, as researchers tend to follow a process that is convenient and proper for the problem they are trying to solve, and usually, the approach is targeted at a specific part of the entire process. The process model by [5] is supposed to address the end-to-end process; hence, it is more elaborate with up to 10 steps. However, most, if not all, of the steps in that process are covered by the consolidated model by Salah et al. Most recently, [20] proposed a process model that was specific for the identification of multistep attacks. This process model had three main steps (1) alert clustering (2) context supplementation, and (3) attack interconnection. This process model assumed all other necessary and preceding steps such as feature selection, normalization, and fusion had already occurred.

#### 1.2.4. Correlation and Clustering Algorithms

Clustering algorithms can be categorized in several ways. One way is whether the algorithm is optimized for numerical or categorical data. Depending on the features selected, network events can be clustered using either category. For example, [21,22] used a numerical-based approach, by calculating the entropy of the different IP header fields to cluster the data. By analyzing the entropy values, certain attack types could be reasoned. As an example, during the scan phase of an attack, few attacking IP addresses attack many ports on a few targets. The entropy value of the destination port addresses would be low for a port-scan attack.

In terms of the categorical features of a network event, researchers have often turned to clustering using the CLIQUE [23] algorithm and its variants. According to the recent comprehensive survey on clustering algorithms performed by Xu and Wunsch [24], CLIQUE can handle high-dimensional data, with a computational complexity that is quadratic with the data dimensions and linear with the number of objects. This means that algorithm performance suffers with high data volumes. The feature set of a network event defined in our paper is five-dimensional; therefore, the algorithm should perform well in a deterministic way. The volume, however, is huge, as network traces from packet sniffers can grow very huge in a small amount of time. Additionally, the first step of CLIQUE is Apriori-like, in which it builds singleton clusters, i.e., clusters of one frequent itemset. It then gradually combines the clusters into bigger partitions of increasing combinations of the feature space before deciding which of those partitions forms a cluster. While this algorithm has successfully been used to identify clusters, the computational complexity of mining frequent itemsets is often cited as its biggest weakness [25].

In terms of mining patterns from event logs, Vaarandi [26] proposed an algorithm that performed data summaries in the first run in the same manner as CLIQUE. In the second step, candidate clusters were created in the cluster table, and the final round of data points were allocated to a matching cluster candidate if it existed; otherwise, a new entry was created. This algorithm was suitable for event logs of textual data, where the feature space could not be figured out beforehand. For real-time clustering, Ma et al. [16] initiated a memory queue Q, into which they deposited an incoming stream of alerts. All alerts that matched a set of mined features were fused into a hyperalert which had features that were a combination of the features of the constituent elementary alerts. They used the temporal similarity to purge a hyperalert from the queue if the incoming alert exceeded a given temporal threshold. The hyperalerts were further stored into a database for further offline analytics. Their approach was tailored to the real-time clustering of alerts and extracted features that were present in typical alert messages but not in the raw events, as is the case with this research.

Our paper concurs with the conclusion by Xu and Wunsch [24] that no clustering algorithm is universally suitable for all problems. With that in mind, we present our own developed correlation and clustering algorithm that considers the uniqueness of the problem at hand. The clustering algorithm takes into consideration the characteristics of network event data to be clustered. Some of the characteristics are summarized below.

- The data space dimensionality is small and fixed. More precisely, the event log entry as defined is a five-tuple record.
- Each event can belong to candidate clusters that can be defined by the set of predefined mining rules.
- The data of the mined features are categorical.

#### 1.2.5. Graph Correlation

Graph correlation presents the alert sequences in an acyclic graph for ease of visualization. The nodes of the graph represent the alerts, while the vertices indicate the temporal relationship between the alerts. Events are modeled as a graph $G = (V, E)$ where the vertex $V$ represents a set of nodes, and each node $(v \in V)$ represents the low-level n-tuple event with $n$ mined features or attributes. The edge $e = (vi, vj) \in E$ represents the connection between the two nodes $v_i$, $v_j$, showing that there is a correlation between the nodes $v_i$ and $v_j$. Typically, this relationship also shows that event $v_i$ precedes event $v_j$. The graph is weighted, where the weight of each edge corresponds to the correlation strength between the event nodes. Consider the events recorded in Table 1.

**Table 1.** Sample network event log entries.

| Event ID | Source IP | Destination IP | Port | Timestamp |
|---|---|---|---|---|
| 1 | 192.168.202.103 | 192.168.207.4 | 53 | 2012-03-16 12:40:35 |
| 2 | 192.168.202.89 | 192.168.207.4 | 53 | 2012-03-16 12:40:41 |
| 3 | 192.168.202.95 | 192.168.207.4 | 53 | 2012-03-16 12:40:48 |
| 4 | 192.168.202.61 | 192.168.207.4 | 53 | 2012-03-16 12:41:25 |
| 5 | 192.168.202.89 | 192.168.207.4 | 53 | 2012-03-16 12:40:41 |

Figure 1 below illustrates the resulting correlation graph constructed from Table 1. The nodes are labeled with the event IDs for ease of reference. The edge weights correspond to the number of features that have matched the similarity rule, while the arrow direction shows the temporal sequence of the correlated events.

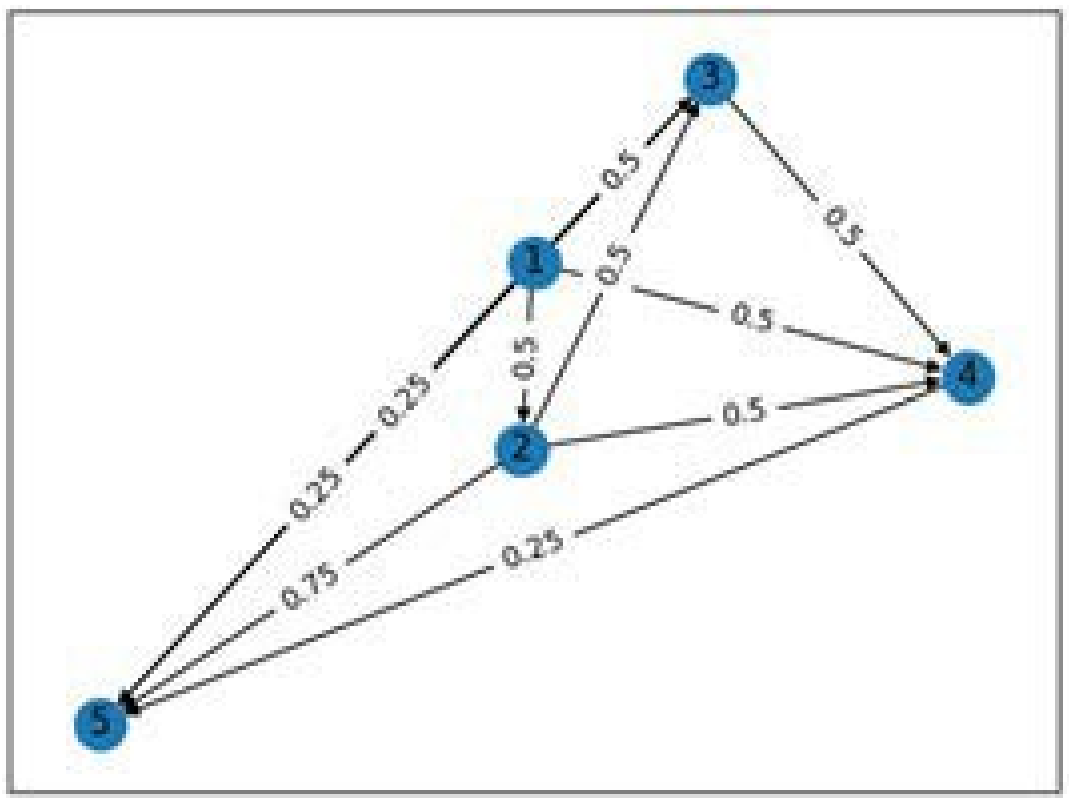

**Figure 1.** Event correlation graph.

Graphs have been used in prior research to construct attack scenarios from intrusion alerts. Some of the works include showing causal relationships among alerts [27], and uncovering multistep attack scenarios [28]. Proximity graphs are used in [29] to produce data points atop of which a modified PageRank algorithm is used to detect anomalies. A vehicle security model is proposed in [30] that uses graph neural network (GNN). The model has spatial and temporal components, and it tries to bring a semantic dimension to the graph correlation to improve the correlation efficiency.

### *1.3. Drawbacks and Limitations*

Firstly, we observe that most works in correlation research relates to alerts. While this has proved to be beneficial, alert correlation is performed outside of the IDS by third-party systems and tools. The IDS itself continues to raise a high volume of alerts, even though the general architecture of an IDS includes a correlation unit [31]. This is seen in all the correlation process models where the input is a stream of IDS alerts. Most organizations do not have automated tools or third-party systems such as SIEM systems to correlate the alerts, and it remains the daunting task of the security analyst to manually sift through them and filter out redundant and irrelevant alerts. This manual process is tedious, time-consuming, and error-prone. Secondly, in the reviewed research on hierarchical correlation, we observe two prevailing approaches. In the first one, as in the case of [6], the same correlation technique is used at both levels of the hierarchy. For example, in this case they used pre–post conditions for both levels. Using the same correlation approach at different levels of hierarchy might enhance the overall correlation, without necessarily eliminating its weakness. The second approach is where an external system is used together with a custom automated tool, such as in [8], where they rely on a third-party system, Splunk, to do the aggregation. This increases operational complexity, as the security team must master the operations of the IDS, the SIEM system (Splunk), and the custom correlation tool, just to accomplish the seemingly simple task of alert analysis. In terms of real-time correlation and detection solutions, we observe that the real-time component only starts from the correlation part of the processes. Cybersecurity resilience requires a process that addresses the requirements of real-time analytics end to end, starting from the time the event is generated, through correlation, to detection. Additionally, most of the works follow the lambda architecture [13], directly or indirectly, where an offline component builds up a historical data set and correlations, and then these are used for the correlations of new alerts. We find that this approach is not the best fit for pure real-time analytics as the time needed to build a good and representative offline component is quite significant. The challenge with in-memory processing is the volatility of RAM, which may result in the loss of the event data. In-memory databases, while they achieve the access throughput requirements of a real-time application, may not be a scalable solution in a direct data acquisition architecture as it would require every event source to have the database. Other options such as caching also have the same volatility challenge. Further, the cache does not always result in a hit.

### *1.4. Research Contribution*

It is no longer a question of whether an organization will be attacked or not. It is rather a question of when and how severe the damage will be. Organizations are therefore forever seeking solutions that can enhance their business continuity in the face of adversarial actions. Timely detection and response are key in today's hostile cyberenvironment. The high volume of alerts is counterproductive to the efforts of security analysts in trying to speedily triage alerts, prioritize them, and execute response actions. Complexity is a security antidote that ought to be avoided by implementing streamlined security operations' processes and systems. This research fills the gap in the current intrusion alert correlation approaches as it aims to address the following issues.

- Reduce the actual volume of alerts generated by an IDS, by correlating the raw events themselves, rather than correlate alerts that have already been raised. This is significant because it places the task of reducing the volume of alerts within the IDS itself. Further, the IDS will raise higher-quality alerts since it will be analyzing correlated information. Further, the IDS can detect more sophisticated attacks as it does not process and analyze data packet by packet as is the case with the state of the art [32].
- Reduce the time taken to respond to threats, because of the reduced number of alerts that need to be analyzed.
- Enhance the capability of intrusion detection systems by improving their correlation unit using a hierarchy of correlation techniques that takes advantage of their cumulative strengths.

The contributions arising out of this research are the following.

1. An event correlation process: All correlation processes reviewed relate to intrusion alerts. Some of the process steps are not relevant to the correlation of raw events. The process model proposed in this research assembles only those process steps that relate to event correlation.
2. A hierarchical event correlation model: The model concentrates on the correlation component of an intrusion detection system and combines similarity and graph-based techniques in an ensemble. The first level of the hierarchy performs aggregation, data reduction, and clustering so that the events ingested by the graph correlation algorithm are significantly reduced. The graph correlation then performs cluster interconnections and visualization to reveal communication patterns and visual analytics in real time.
3. A clustering algorithm, after considering options for clustering the network event data and their unique characteristics, with a time complexity of $\mathsf{O}n$.

The outcome of the research is a streamlined, end-to-end event correlation model that enables real-time attack detection.

## 2. Proposed Approach

Our approach is based on a hierarchical correlation of events. In the reviewed literature, some works have used similar approaches but with fundamental differences from our work. In the table below, we highlight some of the works worth noting.

**Table 2.** Summary of related works.

| Re | Data Source | Technique Used | Approach |
|---|---|---|---|
| Patton et al. [8] | Network alerts | (1) Splunk is used to aggregate alerts<br>(2) Text analysis of the created hyperalerts | **Splunk** is used to perform the first level of the correlation to aggregate IDS alerts. The aggregated alerts are then further clustered using text analysis. |
| Tian et al. [7] | Network alert | (1) Attribute similarity<br>(2)Prerequisite and consequences | IDS alerts are initially correlated using attribute similarity to create hyperalerts. The resulting hyperalerts are then correlated using prerequisites and consequences. |
| Cuppens and Miege [6] | Network alerts | Pre–post conditions | IDS alerts are correlated using pre–post conditions to discover and generate hyperalerts and discover local attacks. The same technique is applied again on the hyperalerts to discover global attacks. |
| Ning et al. [33] | Network alert | (1) Attribute similarity<br>(2) Prerequisite and consequences | IDS alerts are initially correlated using attribute similarity to create hyperalerts. The resulting hyperalerts are then correlated using prerequisites and consequences. |
| Diakhame et al. [9] | IDS | NLP using NER and semantic similarity | Attributes of aggregated alerts are extracted using NER and then correlated based on semantic similarity. |
| Proposed Model | Network events | (1) Attribute similarity and (2) graph correlation based on time and association rules | Raw event records are initially correlated based on attribute similarity to achieve aggregation and create hyperalerts. The hyperalerts are then correlated using both temporal features and association rules to establish global traffic patterns that may flag suspicious flows. |

The fundamental difference between all these works from the approach proposed in this research is that they perform correlation after intrusion detection by sensors such as IDSs. While this prevailing approach will still reduce the alerts in the end, the approach taken in this research is to correlate before the alerts are produced by the sensor. Incidental to the above fact, all the reviewed systems operate on alerts, while in this approach the correlation is performed on raw events. The significance of this is that the volume of alerts is reduced at the IDS level itself since the IDS operates on a reduced data set. Additionally, the IDS can now detect some complex and multistep attacks since it analyzes already correlated hyperevents. This capability is not possible when analyzing singular events. This is particularly beneficial for security operations that do not have SIEM systems and other third-party correlation systems since the correlation component becomes integral to the IDS itself. The other difference is in the techniques used and how they affect the overall performance of the system. Both [7] and [33] use attribute-based correlation as a first stage to produce hyperalerts, then they cluster the hyperalerts further by correlating them using pre–post conditions. Pre–post conditions require a knowledge base of such rules to capture every possible attack. If the pre–post conditions do not cover all possible attack scenarios, unknown attacks cannot be discovered. In the case of the research by Cuppens and Miege [6], they also used pre–post conditions; however, they used the same technique at both levels of the hierarchy. Using the same correlation technique at different levels of a hierarchy might enhance the strength of the technique without necessarily eliminating its weaknesses. In the case of [8], they rely on a third-party system, Splunk, to do the aggregation and then use a custom automated tool to cluster the aggregated alerts using text analysis. This approach increases operational complexity as the security team must master the operations of the IDS, the SIEM system (Splunk), and the custom correlation tool. Operational complexity is undesirable, even more so in information security, where pragmatism, nimbleness, and agility must be the operational ethos. In addition to the above differences, our approach offers the benefit of visualization due to the graph correlation at the second level of the hierarchy [38,39,41].

## 2.1. *Event Correlation Process*

The proposed process model is motivated by the alert correlation processes in [20] and [43,44] and optimizes them by considering only those steps that are relevant for event correlation. The proposed process is shown in Figure 2 below.

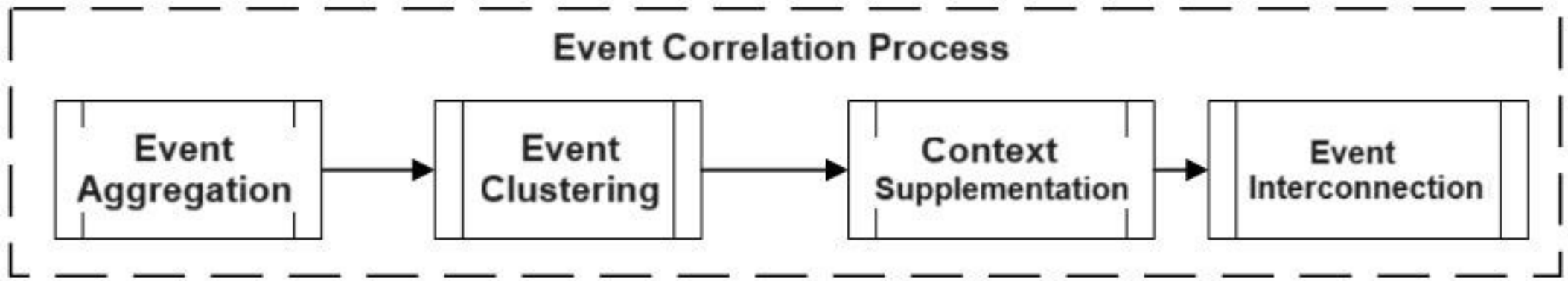

**Figure 2.** Event correlation process.

The details of the process model are covered in the context of the correlation model presented next.

## 2.2. *Hierarchical Event Correlation Model*

The hierarchical event correlation model comprises two levels of correlation. The first level performs the aggregation, clustering, and context supplementation steps of the process model, while the second level performs event interconnection and additional context supplementation [47]. The input to the model is a stream of raw events such as network event records from a packet sniffer or live digital feeds of intelligence data such as vulnerability feeds or blacklisted IP addresses and domains. The output from the first level are clusters of events representing network sessions between pairs of communicating hosts at varying levels of correlation. These clusters become the input to the second level. The second level further clusters these events and merges them to create higher-level representations of the communication flows. The output from the second level is communication graphs which can be analyzed visually or by an IDS to uncover suspicious patterns that may indicate an attack [46]. The high-level view of the proposed event correlation model is shown in Figure 3 below.

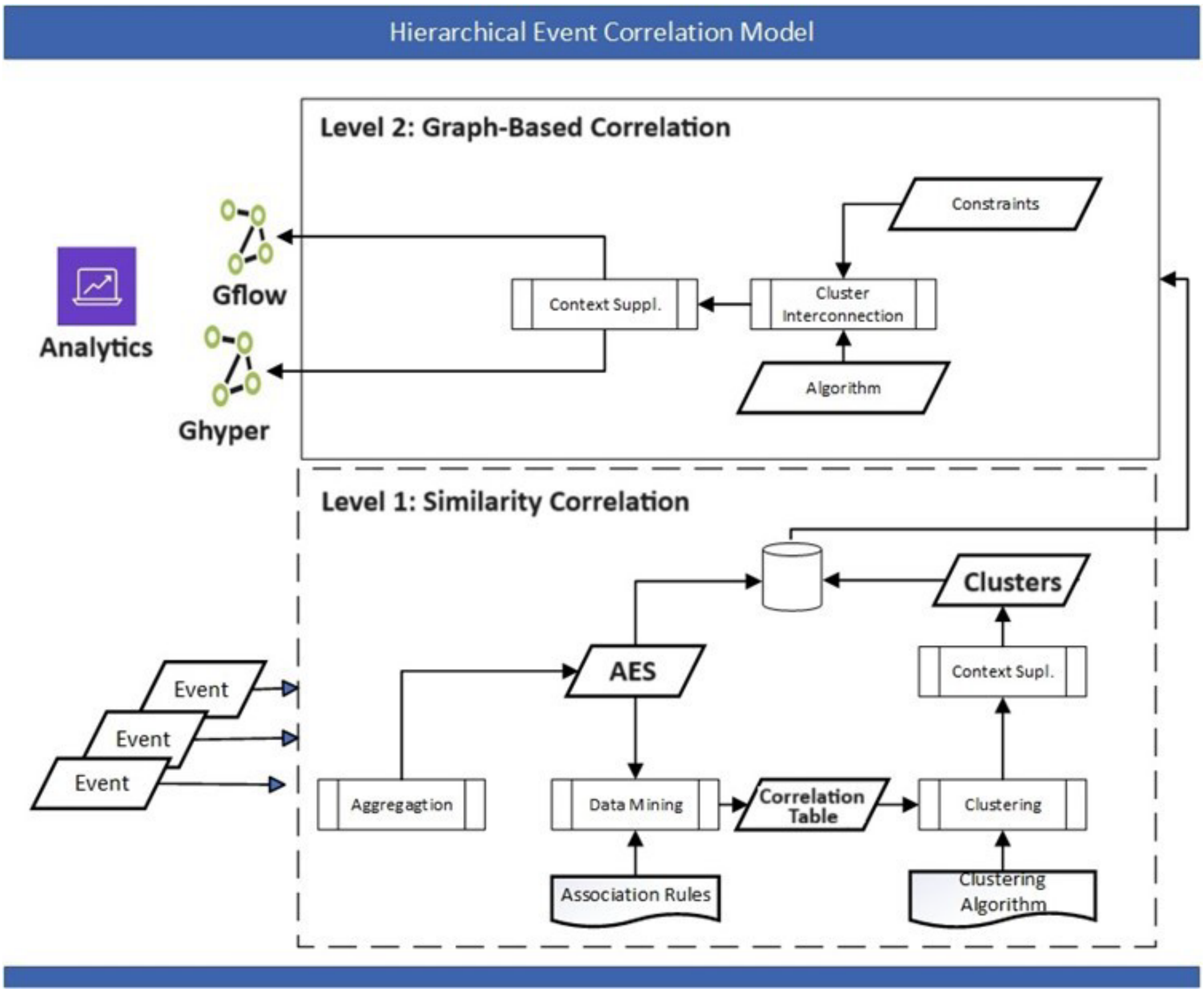

**Figure 3.** Hierarchical event correlation model.

The details of each step of the process are covered in the subsequent sections for each level of the correlation hierarchy.

## *2.3. Level 1 — Similarity-Based Correlation*

The first level is aimed at the aggregation, clustering, and context supplementation steps of the event correlation process defined. This model uses the similarity of event attributes for the correlation. The input to the first level is the stream of events. The output is a set of clusters that have been processed and have had contextual information added and carried through the process.

### 2.3.1. Event Aggregation

The purpose of aggregation is to summarize events that convey the same knowledge within some defined observation window so that one aggregate event is sufficient to represent all the transactions recorded in the individual events. This model defines an event as an n-tuple $e = \langle attr_1, attr_2, \ldots, attr_n \rangle$, where each $attr_i \in e$ represents an extracted attribute or feature out of all the possible features of the event. The aggregated event E is the fusion of all such events having all similar features, except the timestamp. Since one aggregate event can represent several individual events, the result of aggregation is fewer events with no loss of information. Which features to mine to form the event depend on the attacks that need to be detected by the model.

In the sample entries in Table 3, all events record the same fact, that is, the host 192.168.204.69 is accessing the service on UDP port 67 (DHCP) running on the target host 192.168.204.1. The only difference in these events is the timestamp. Ignoring the timestamp, these events can be represented by the aggregate event E shown in Table 4 below.

**Table 3.** Individual event log entries.

| Timestamp | Src_IP | Dst_IP | Src_Port | Dst_Port | Protocol | Bytes |
|---|---|---|---|---|---|---|
| 12:31:57.740 | 192.168.204.69 | 192.168.204.1 | 68 | 67 | UDP | 328 |
| 12:33:02.540 | 192.168.204.69 | 192.168.204.1 | 68 | 67 | UDP | 328 |
| 12:35:00.750 | 192.168.204.69 | 192.168.204.1 | 68 | 67 | UDP | 328 |
| 12:38:03.760 | 192.168.204.69 | 192.168.204.1 | 68 | 67 | UDP | 328 |

**Table 4.** Aggregated event, E.

| Timestamp | Src_IP | Dst_IP | Src_Port | Dst_Port | Proto | Bytes |
|---|---|---|---|---|---|---|
| 12:31:57.740 | 192.168.204.69 | 192.168.204.1 | 68 | 67 | UDP | 1312 |

In the aggregate event, all the features are the same, except the number of bytes (bytes), which is extracted from the length field of the IP header. The count and packets (PKTS) are calculated fields which track the number of packets that make up the aggregate event. To maintain the information from the elementary events in the aggregate, the bytes attribute of the aggregate is the total of the individual byte fields from the constituent events. This is done so that there is no loss of information from the individual events. In this way, analysis and detection can be based on this aggregate in the same way it would have, based on the individual events. The output of the aggregation step is an aggregated event set *AES* comprising a set of the aggregate events *E* as defined by the following equation.

$$AES = \{E_1, E_2, \ldots, E_n\} \tag{1}$$

2.3.2. Event Clustering

Before the events can be clustered, association rules are defined to mine data from the *AES* that represent the fundamental analytics for network communications. Each aggregate event *E* in *AES* has *n* candidate correlations, where n is the number of features that make up the aggregate event. To mine the correlation candidates, a minimum of n association rules is created for each candidate cluster. The general form of a rule R is given below.

$$R = \{attr_1, attr_2, \ldots, attr_k\},\ sup \geq k \leq n \tag{2}$$

A support value sup dictates the minimum frequent itemsets to be considered for the correlation, while n is the number of all the features of the aggregated event. This allows the creation of candidate correlations at varying levels of correlations with different values of k. In the implementation of this model, the network event log entry is defined by the 5-tuple below.

$$Net_L = \{S_{IP}, S_{port}, D_{IP}, D_{port}, Protocol\} \tag{3}$$

Based on the network event log, the mining association rules below are defined.

$$SC : A_i = B_i\ \forall i \in \{S_{IP}, D_{IP}, S_{port}, D_{port}, Protocol\} \tag{4}$$

$$VC_1 : A_i \equiv B_i\ \forall i \in \{S_{IP}, D_{IP}, S_{port}, D_{port}\} \tag{5}$$

$$VC_2 : A_i \equiv B_i\ \forall i \in \{S_{IP}, D_{IP}, S_{port}\} \tag{6}$$

$$VC_3 : A_i \equiv B_i\ \forall i \in \{S_{IP}, D_{IP}, D_{port}\} \tag{7}$$

$$VC_4 : A_i \equiv B_i\ \forall i \in \{S_{IP}, D_{IP}\} \tag{8}$$

Shared correlations, SC, indicate that the same host is communicating on the same applications. Correlation with variance level 1, VC1, is achieved when the same source host talks to the same target host on the same application but using different protocols. Correlation with variance level 2, VC2, is when the same host and the same target are communicating on the same application using the same source ports, while the destination ports or protocol could be different [45]. In VC3, the same source is accessing the same target host and destination port but could be using different source ports. When only the source and destination IP addresses are similar, this is mined by the VC4 rule. A careful inspection of these rules should reveal that they represent the foundational analytics for analyzing network events. Other types of analytics can be derived from various correlations or interconnections of the clusters created. For example, the SC represent normal traffic in general terms. However, if the number of these correlations and the volume of traffic keeps increasing in a time series analysis, there could be a DoS attack in flight. Similarly, an increase in VC2 correlations could signal a probe attack, as the source is sending so many packets to the destination from different source ports. The data mined by these rules are put into a correlation table, *CorrTable*, which is an in-memory hash table of candidate clusters for each aggregate event that satisfies the minimum support, sup. This correlation table is passed onto the clustering algorithm which allocates the data points by picking the best-fitting cluster of the candidate clusters for each event. Event clustering groups similar events into cluster C such that each cluster has some contextual meaning using some clustering algorithm. Typically, a measure of similarity is defined which is used to decide whether two events should be put in the same cluster. The idea is that events within the same cluster should generally be given the same analytic interpretation or detection treatment. Different criteria can be used to contextualize the clusters for network events, such as clusters of source addresses, destination addresses, applications, and others. The output from this stage is a set of clusters $C = \{C_1, C_2, \ldots, C_n\}$ such that each cluster $C_i \in C$ has a unique semantic relevance. In an implementation of this model, the clustering is performed in a custom-developed clustering algorithm, which is shown by the pseudocode in Figure 4 below.

```
Function Cluster
Input net_log, support,
corr_table = {}, cluster_table
Rules = {SC, VC1, VC2, VC3, VC4}
For event in net_log:
    data = {}
    For feature in rule:
        data[feature] = event[feature]
        create a hash for the mined data
        look up the hash in corr_table
        if the result set is empty:
            create a new meta_event
            count = 1
            assign label in {SC, VC1, VC2, VC3, VC4}
            pkts = event["pkts"]
            bytes = event["bytes"]
            ts = event["ts"]
            insert the meta_event into corr_table using the hash as the key
        else:
            count += 1
            pkts += event["pkts"]
            bytes += event["bytes"]
            if ts > event["ts"]
            ts = event["ts"]
    For all meta_events in corr_table:
        if count > support:
        insert the meta_event into the cluster table
        insert the meta_event into the database
Output cluster_table
```

**Figure 4.** Event clustering algorithm.

The algorithm takes the batched-up event streams as input. Hash tables for the correlation and cluster tables are initialized on line 3. Lines 5–8 mine data from the record based on the current rule index. The hash for the current rule data is created in line 9 and used to perform a lookup on line 10. From lines 10 to 18, a new metaevent is created if the lookup returns no match. The metaevent is then inserted into the correlation table on line 18. If there is a match, lines 19–24 update the metaevent. The cluster table is a subset of the correlation table for all metaevents whose count satisfies the support. This is indicated in lines 24–28. The output of the function is the cluster table on line 29.

#### 2.3.3. Context Supplementation

Context supplementation is used to provide some descriptive information to the clusters to help analysts better understand what the cluster is representing. The type of information that can be used is flexible, depending on the use case of the cluster. In this model, context supplementation is performed at both levels of the hierarchical correlation. During the first phase, at a minimum, each cluster is given two descriptive attributes: (1) count and (2) label. A count is a number that indicates how many elementary events were fused to create the metaevent, while the label $l \in L =$☯$SC, VC1, VC2, VC3, VC4$☸. The interpretation and significance of these labels correspond to the descriptions provided by the corresponding mining rules given in the clustering section above. This allows analytics to be performed for flows of a given label.

### 2.4. *Level 2 – Graph Correlation*

The clusters from the first level are fed into the second level for additional correlation using graphs. This level establishes cluster interconnections and communication patterns and enables visual analytics. The input is the set of clusters of events from the level 1 correlation above, while the output is a series of interconnections and communication graph flows. Figure 5 below details the level 2 correlation block.

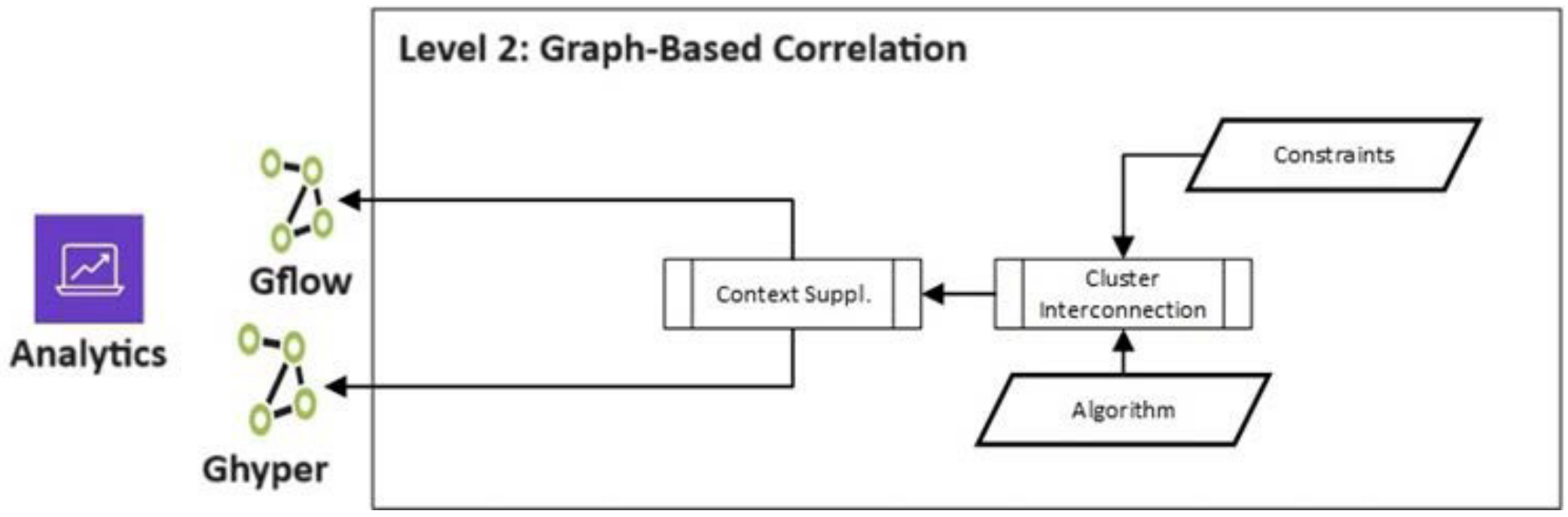

**Figure 5.** Graph-based correlation.

#### 2.4.1. Cluster Interconnection

The cluster interconnection stage takes the clusters created in the previous step and finds similarities between them to create a bigger picture view of the underlying communications captured in the clusters. The interconnection stage is used to reveal communication patterns in the network to establish normal and suspicious patterns. This model defines two graphs that are created to reveal cluster interconnections.

**Definition 1.** *A hyperevent correlation graph is defined as Ghyper* 🖬 ☯*M, E*☸*, a sequential, directed acyclic graph (DAG) representing correlations among clusters of events. The set M of nodes is the set of metaevents for all communication flows between the same source and destination hosts within the observation window. The metaevents are the clusters formed during the level 1 correlation described above. For two nodes $m1, m2 \in M$, an edge $e \in E$* 🖬 ☎*$m1, m2$*☽ *is created that connects $m1$ to $m2$ if the source and destination are similar for both metaevents. What constitutes a source and a destination is left to the implementation, according to the features mined. Possible examples* *could be source and destination IP addresses, source and destination port numbers, source and destination mac addresses, or any combination of source and destination features. Additionally, the edge is formed if metaevent* $m1$ *occurred before metaevent* $m2$ *according to the timestamps for the events. By sequencing the events, analytics that are based on the sequence of events can be executed by examining the data structure of Ghyper. In general, an edge in Ghyper is created if all the conditions below are satisfied.*

$$m1.src == m2.src \tag{9}$$

$$m1.dst == m2.dst \tag{10}$$

$$m1.ts < m2.ts \tag{11}$$

The objective is to interconnect related clusters and create a set of clusters of the hyperevents $H = \{c_0, c_1, \ldots, c_n\}$, such that each cluster $C_i \in C$ represents a collection of metaevents from the previous step with some similar attributes. This establishes multisession communication correlations among the single-session clusters established in stage 1 for the same communication pair. *Ghyper* is the first interconnection step of the process. Additionally, this model defines the graph $Gflow$ which shows all the communication flow patterns in the network.

**Definition 2.** *A communications flow graph, GFlow* $= (M, E)$*, is a directed acyclic graph (DAG) that shows the flow of communications in the network during the observation window. The set M of nodes is the set of metaevents from the level 1 correlation. For two nodes* $m1, m2 \in M$*, an edge*
$e \in E = \{m1, m2\}$ *is created that connects* $m1$ *to* $m2$ *if the destination of* $m1$ *is the source* $m2$*. An*
*edge in Gflow is formed if the condition in the equation below is satisfied. Just as in Ghyper, what constitutes a source and a destination is left to the implementation.*

$$m1.dst == m2.src \tag{12}$$

The principle is that an edge is formed if traffic is flowing from m1 to m2. Consider the communications in Table 5 below.

**Table 5.** Communication flows.

| Source | Destination |
|---|---|
| A | B |
| A | C |
| B | C |

The above communications could be represented by the following Gflow shown in Figure 6

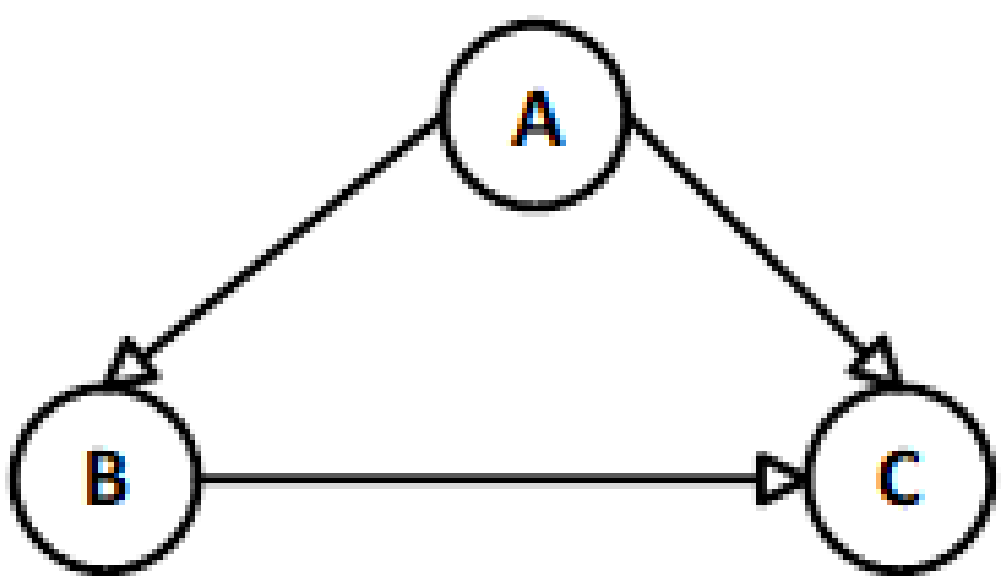

**Figure 6.** Conceptual Gflow.

Gflow provides important analytics. A node with multiple outgoing edges, for example, node A in the above graph, could indicate a high talker, or an intruder could be carrying out attacks such as sweep probes on hosts B and C. A node with multiple incoming edges, such as node C above, could be a popular service, such as DNS or DHCP, but it could also be a node under a DDoS attack. The weakness of graph correlation is that if the data are too big, the run time could be long. Network events are classified as big fast data, owing to their huge volume. If these big data were to be presented to a graph correlation algorithm, the time taken to complete the correlation would be too long and defeat the whole idea of real-time analytics. In the implementation of this model, the source and destination are taken to be the source and destination IP addresses. Equations (9)–(12) are then interpreted by the implementation equations below to implement *Ghyper*. Equation (12) defines a generic source of events, leaving it to the specific implementation to specify what the source is. Possible sources could be source IP addresses, source port numbers, source MAC addresses, or any combination of the various source options. In this experiment, the source was the source IP address of network packets. In the same manner, the destination in this experiment was the destination IP address of network packets. This results in the fine-tuning of Equations (9) and (10) into Equations (13) and (14). Further, Equation (12) was fine-tuned to the following equation. By aggregating and correlating the events from the first level, the resulting input to the graph correlation algorithm is very small and manageable, allowing the algorithm to complete in real time. It can therefore be observed that the two approaches in the correlation hierarchy are complimentary and deliver an outcome that is impossible to achieve with either one of them in isolation, with a performance acceptable for real-time applications. Using these two defined graphs, a variety of analytics can be developed. Most importantly, the detection component of the model can base its analysis on these graph data structures.

## 3. Experiments and Results

### *3.1. Data Sets*

One of the biggest challenges facing intrusion detection research is the availability of realistic data sets. Due to various privacy concerns, organizations are often unwilling to provide real network traffic data for research. Consequently, researchers often use synthetic data sets or data sets created through simulations. The survey by [31] lists and analyzes the various publicly available data sets for intrusion detection research. We used the CICIDS-2017 data set, which is one of the most recent data sets to have emerged for intrusion detection research. The CICIDS-2017 data sets contain labeled flows as an Excel spreadsheet in .csv format. Each record in the spreadsheet is a result of parsing the corresponding PCAP file and extracting about 83 statistical features for each flow. We used data from the afternoon of Friday, 7 July 2017. The afternoon data contain a mixture of labeled benign network traffic flows, various forms of port scans, as well as DDoS LOIT attacks. This experiment extracted one hour’s worth of traffic from the data set.

### *3.2. Environment*

The environment for the experiment was set up as in Figure 7.

A Windows machine was set up as the data collector. The data collector implemented a custom application written in Python 3.11. The parser comprised several functions to parse various formats of source files including .pcap, csv, syslog, and evtx. For this experiment, the CSV file of features and labels was read line by line and a Python dictionary was constructed for each flow record. Each flow dictionary was then immediately streamed using Apache Kafka.

A Linux server machine was set up in the public cloud. The server was equipped with 64GB of RAM, 2TB of hard drive space, and eight processors with two threads per core. On this server, three virtual machines were set up. The first machine implemented the correlation and detection component of the experiments. A Python script was developed as a Kafka consumer to read live events as they arrived at the Kafka server. The events then were passed on to the analytics application, running inside the same virtual machine. The analytics application then performed the correlation. At the end of each correlation iteration, the raw events and the completed analytics were sent to a MongoDB database which was set up in a separate virtual machine. The third virtual machine was running a snort IDS. This was used to process the same original packet capture file to check for intrusions, for comparison with the performance of the developed detector. The table below summarizes the environment setup.

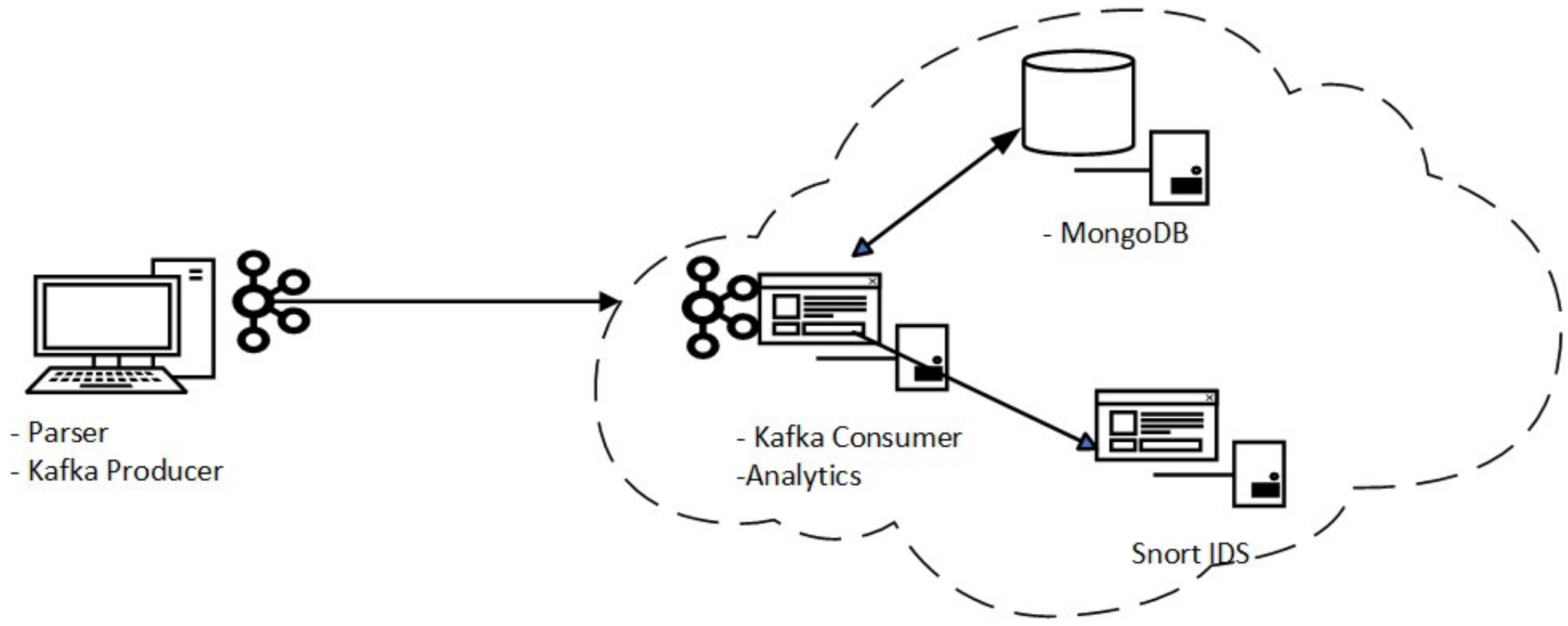

**Figure 7.** Experimental setup.

**Table 6.** Platform configuration for the experiment.

| | Function | Hardware | Virtual Machines |
|---|---|---|---|
| **Client machine** | Data collector | Lenovo IdeaPad S340<br>Intel i7 4 Core Processor<br>8 GB RAM<br>Windows 10 Pro | None |
| **Server** | Cloud server | DELL<br>AMD Ryzen 7 PRO 3700<br>64 GB RAM<br>Ubuntu 22.04 | Corellator app<br>MongoDB<br>Snort IDS |

### 3.3. Results and Discussion

The correlation model developed in this paper firstly aggregates all events into an aggregated events set. The aggregated events are then correlated with different levels of variations forming clusters according to mining rules. The results show that in most cases the shared correlations reduced the data set by smaller percentages, as can be noted by the difference between the number of events and clusters. This is unsurprising because in this data set the flows from the csv data are already aggregated at 1-minute intervals. There are no observed VC1 clusters. VC1 would be formed if two flows only differed by the protocol type, for example, if a host sends a packet to port 53 (DNS) using both TCP and UDP. On the other hand, there are VC3 correlations on all flows. Our analysis of the clusters revealed that these contained either traffic to popular and public services such as DNS and HTTP, or scanning attacks. Table 7 summarizes the results of the first ten time slots of the correlation.

The minimum support was chosen to be one for this experiment. This was done so that probe attacks were captured, as most probes would not correlate at higher levels of support. The distribution of the correlations is plotted in the pie chart in Figure 8. The pie chart shows that the majority of the traffic was for shared correlations (SC). Benign traffic is expected to exhibit this correlation behavior. Inspecting the cluster distribution allows for visual analytics. This becomes important for analytics applications such as intrusion detection systems as analysts can use the visual clues to perform further detailed investigations into suspicious clusters. Different analytics can be performed on the clusters to uncover interesting patterns of communications contained within the data set.

**Table 7.** Correlation summary.

| Time_Slot | Events | AES | VC1 | VC2 | VC3 | VC4 | Meta_Events | Graphs |
|---|---|---|---|---|---|---|---|---|
| 0 | 258 | 228 | 0 | 0 | 25 | 11 | 82 | 31 |
| 1 | 1258 | 1156 | 0 | 0 | 64 | 3 | 206 | 78 |
| 2 | 1558 | 1432 | 0 | 0 | 104 | 6 | 314 | 155 |
| 3 | 1362 | 1261 | 0 | 0 | 86 | 7 | 374 | 118 |
| 4 | 968 | 8960 | 0 | 0 | 79 | 4 | 272 | 114 |
| 5 | 1597 | 1470 | 0 | 0 | 92 | 9 | 361 | 133 |
| 6 | 1225 | 1094 | 0 | 0 | 77 | 2 | 322 | 141 |
| 7 | 1504 | 1354 | 0 | 0 | 97 | 5 | 328 | 128 |
| 8 | 1461 | 1327 | 0 | 0 | 81 | 5 | 311 | 117 |
| 9 | 1582 | 1438 | 0 | 0 | 115 | 5 | 382 | 174 |

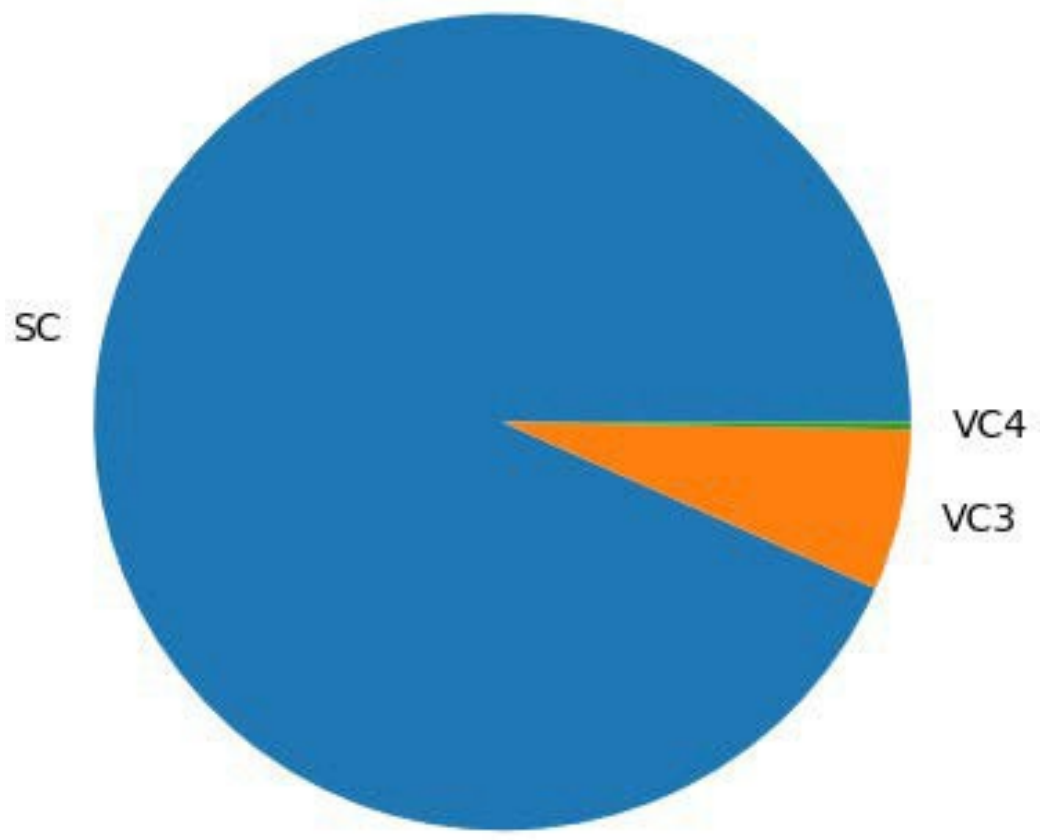

**Figure 8.** Distribution of clusters in the data set.

*Ghyper* interconnects all clusters with the same source and destination IP address pairs. From the clusters described in Table 7, the output of the *Ghyper* correlation is shown in Figure 9 for time slot 1.

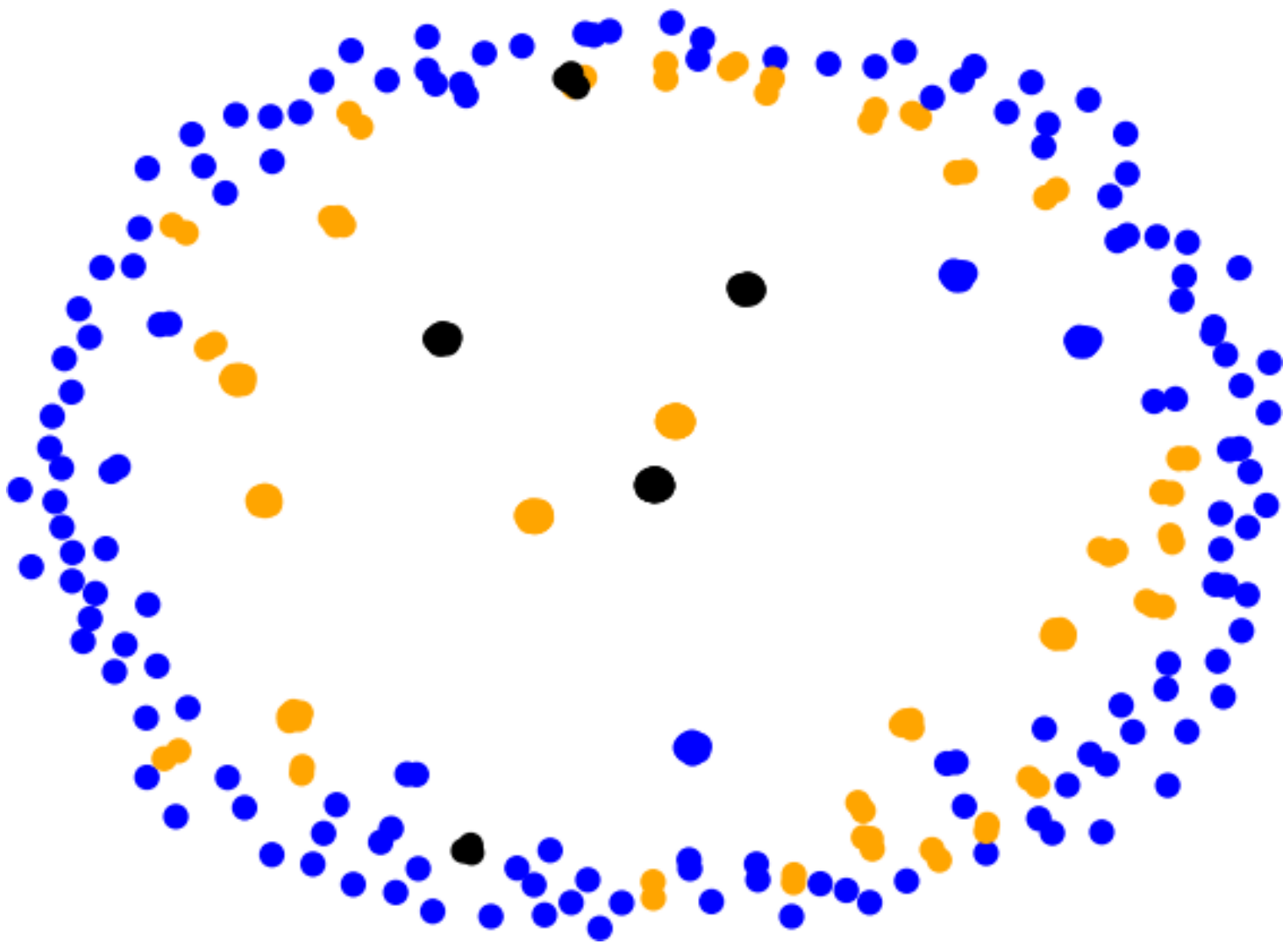

**Figure 9.** *Ghyper*.

Each cluster in the *Gyhper* graph represents a correlation of all VCx metaevents between source and destination communication pairs within the measurement time window. Zooming into each cluster shows the individual metaevents (VCx) that make up the cluster.

In the graph shown, the blue clusters are the shared correlations. The orange and black clusters are VC3 and VC4, respectively. It can be seen from the graph that there are very few VC4 clusters in the data set. VC4 are clusters where the flows have nothing in common except source and destination IP addresses [40] and [42].

To investigate the flow of communication between the pairs, the Gflow analytic was used. Gflow shows the communications paths in the data set from each source to various destinations. The graph in Figure 10 shows Gflow for the data in time slot 1.

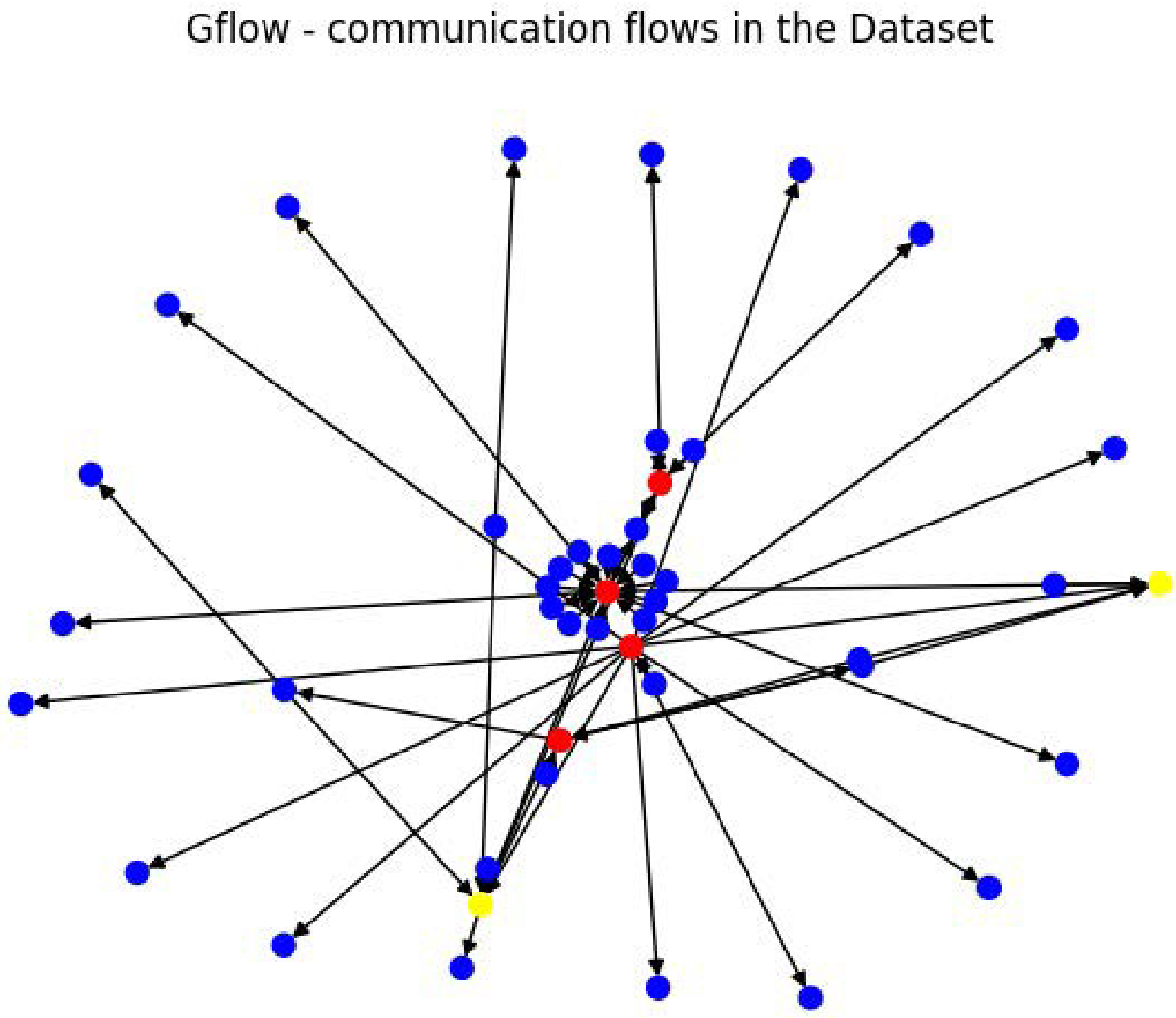

**Figure 10.** Cluster interconnection graph for time slot 1.

The nodes in Gflow represent different communicating sources, identified by their source IP addresses. The edges are created between two communicating hosts, with the direction of the arrow showing the direction of the traffic flow. This analysis can provide a quick visual analysis of the network. Where a node has many outgoing edges, it shows a high talker. Many incoming edges would be indicative of a popular service being available on that host, but it could also signal a potential DDoS attack in progress. One of the appealing elements of graph correlation is visualization. In the graph above, the red color was used to flag out all nodes that had more than five outgoing connections. More outgoing connections could signify a sweep attack if the trace contains probe traffic. This helps the analyst to investigate nodes for further details. The parameter for flagging is configurable in the analytics. According to the Gflow graph above, four hosts had at least two outgoing connections to different targets in the data set under analysis. The orange color is for nodes exceeding a threshold number of incoming connections. In this experiment, the threshold was two.

The first aspect of correlation performance is the resource utilization of the correlation system by measuring the time of the entire correlation process (aggregation, level 1 correlation, clustering, and level 2 graph correlation) as well as peak memory usage. The measurements were taken for each time slot. From the results shown earlier in Table 7, the charts in Figure 11 show the correlation performance.

As can be seen from the charts, the correlation process's run time and peak memory usage increased with an increasing number of events, clusters, and hyperevents.

Apart from the run-time performance, two metrics are commonly used to measure the effectiveness of the correlation. A core component of our model, and all correlation systems, is to aggregate the data source into a smaller data set. Two metrics are used for this measurement. Completeness measures how well the events are correlated and is given by Equation (13).

$$R_c = \frac{correctly\ correlated\ events}{related\ events} \tag{13}$$

Soundness measures how correctly the events are correlated and is given by the equation below.

$$R_s = \frac{correctly\ correlated\ events}{correlated\ events} \tag{14}$$

To measure the performance of the correlation algorithm, the same approach as above was used; however, the time was measured only for the clustering algorithm part of the correlation process. The algorithm run-time performance is plotted in Figure 12 below.

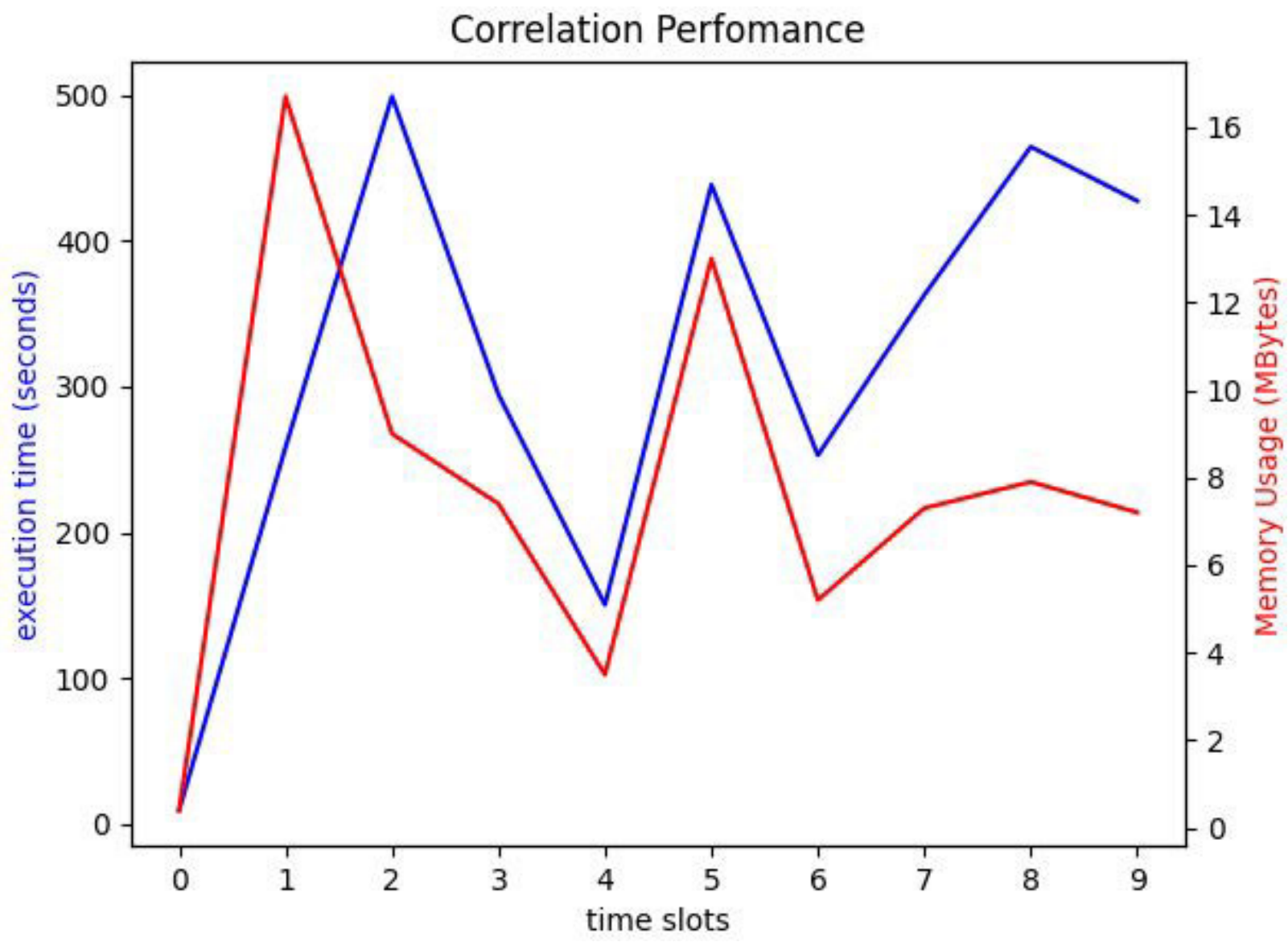

**Figure 11.** Correlation performance.

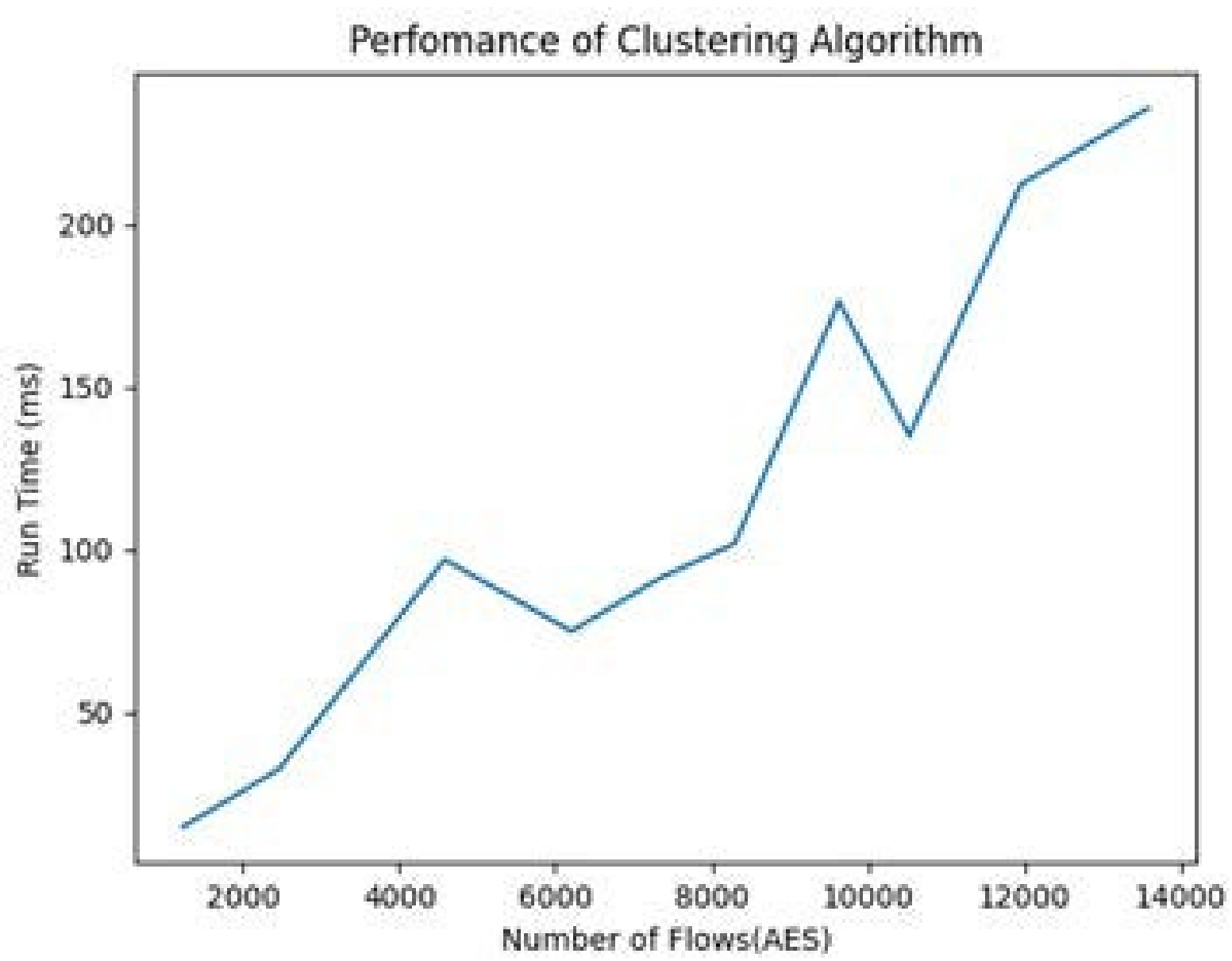

**Figure 12.** Performance of the clustering algorithm.

The results show that in general, the run time of the algorithm increased linearly with the size of the input. For an input of size n, the performance observed was ***O(n)***.

Our clustering algorithm has a time complexity of $O(n)$. This is better performance compared to the complexity of alternatives such as K-means and CLIQUE, which have time complexities of $O(n^2)$ [34] and $O(3^{n/3})$ [35], respectively, where $n$ is the input data size.

Table 8 compares the correlation performance of our model with the work of other related works.

**Table 8.** Comparison with related works.

| Ref. | Approach | Dataset | RC | RS |
|---|---|---|---|---|
| Lin et al. [11] | The correlation is based on prerequisite and consequence with real-time data collection based on software agents. | CERCNET2 TESTBED | 0.77 | 0.81 |
| Wang et al. [36] | They propose a correlation system that uses alert semantics (based on the description field) and attack stages. | DARPA LLDOS 1.0 | 0.92 | - |
| Anbarestani et al. [37] | They use Bayesian network-based alert correlation to discover attack strategies | DARPA LLDOS2.0.2 | 0.91 | 0.72 |
| Diakhame et al. [9] | They develop a correlation framework for cyberattack reconstruction utilizing natural language processing (NLP) and semantic similarity techniques | UNSW-NB15 | 0.54 | - |
| Our model | The correlation of raw events is performed using a hierarchy of (1) temporal and attribute similarity and (2) graph-based correlation for uncovering suspicious network communication patterns | CICIDS-2017 | 0.87 | 1 |

## 4. Conclusions

This paper explored a novel approach to solving the problem of the high volume of alerts raised by intrusion detection systems. The main contribution is a hierarchical event correlation model that ingests raw events from the source and outputs correlated graph data structures to the detection part of the IDS. We showed that by correlating the raw events rather than alerts, the system reduced the complexity of security operations since it could be integrated into an IDS as a core correlation component; this contrasts with other approaches seen in prior art which rely on third-party systems to correlate alerts that have already been raised. We see an implementation of this model being integrated into a distributed IDS, where the data collection, correlation, and detection components are spread into different nodes. We showed that no intrusion information was lost during the correlation process. The input to the detection component was a significantly reduced set of aggregated events. Based on this observation, the IDS should detect threats just as it would based on the elementary events. This approach enables the IDS itself to raise fewer alerts, which enables a fast and real-time response to threats. The proposed model proves that by combining two complementary correlation techniques, the overall system is more efficient and effective. Our model employs similarity and graph-based correlation techniques in a hierarchy.

The similarity-based correlation is performed at the first level of the hierarchy to achieve aggregation, data reduction, and clustering. Graph-based correlation is performed at the second level to interconnect related clusters and communication patterns and provide event visualization for ease of analysis. By combining these two approaches, the system benefits from their complementary strengths to achieve an integrated capability. Additionally, this research contributed to the security event correlation process. The process was used in the development of the proposed model to show its applicability. It is a process that is tailored to the correlation of raw events rather than alerts. In the future, we intend to develop an integrated correlation-based intrusion detection component as a proof of concept to consume the correlated data, analyze, and detect threats. This work is in progress and already at an advanced level.

**Author Contributions:** Conceptualization, H.M. and K.O.; methodology, H.M., K.O., and M.C.G.; software, H.M. and K.O.; validation, H.M., K.O., and M.C.G.; formal analysis, H.M., K.O., and M.C.G.; investigation, H.M., K.O., and M.C.G.; resources, K.O.; writing—original draft preparation, H.M., K.O., and M.C.G.; writing—review and editing, K.O. and M.C.G.; supervision, K.O. and M.C.G.; project administration, H.M. All authors have read and agreed to the published version of the manuscript.

**Funding:** This research received no external funding.

**Informed Consent Statement:** Not applicable.

**Data Availability Statement:** Data and codes are available upon request.

**Conflicts of Interest:** The authors declare that they have no known competing interests or personal relationships that could have appeared to influence the work reported in this paper.